\documentclass[onecolumn,showpacs,aps,amssymb,floatfix,prd,amsmath,preprintnumbers]{revtex4}
\usepackage{epstopdf}
\usepackage{capt-of}
\usepackage{graphicx}  
\usepackage{dcolumn}   
\usepackage{float}
\usepackage{hyperref}
\usepackage{subfig}

\begin{document}
\title{Oscillatons in Scalar-Field Dark Matter from a Full Fourier Expansion of an Exponential Potential}

\author{ A. Mahmoodzadeh $^{1}$\footnote{Corresponding Author, E-mail: Ali.Mahmoodzadeh@iau.ac.ir}, K. Ghaderi $^{2}$\footnote{ E-mail: k.ghaderi@iau.ac.ir}. P. Amiri $^{3}$\footnote{ E-mail: amiriprshang@gmail.com}}

\affiliation{$^{1}$ Department of Physics, Bou.C., Islamic Azad University, Boukan, Iran.}
\affiliation{$^{1}$ Department of Physics, Mari.C., Islamic Azad University, Marivan, Iran.}
\affiliation{$^{3}$ Department of Physics, Bou.C., Islamic Azad University, Boukan, Iran.}

\begin{abstract}
Real, time-dependent scalar fields can form oscillating, self-gravitating configurations-\emph{oscillatons}-that are viable candidates for scalar-field dark matter (SFDM). We revisit oscillatons with an \emph{exponential} self-interaction and develop a full Fourier (Jacobi–Anger) treatment that resums the time dependence of both the metric and the potential, thereby unifying quadratic, quartic, and higher-order interactions within a single framework. After fixing the small-amplitude normalization \(V_{0}=m_{\Phi}^{2}/(\lambda^{2}k_{0})\), we derive a closed, dimensionless boundary-value problem for the radial profiles and solve it numerically via Bessel-series truncation with controlled convergence. We compute time-resolved and time-averaged observables energy density, radial energy flux, radial/tangential pressures, and total mass and map their dependence on the coupling \(\lambda\) and central amplitude. The geometry exhibits only even harmonics of the fundamental frequency, while composite observables inherit a DC part plus even harmonics; the radial flux oscillates predominantly at \(2\omega\). Apparent negative instantaneous pressures arise from coherent oscillations and are assessed consistently through classical energy-condition diagnostics (WEC/NEC/SEC). Our formulation provides a reproducible and extensible baseline for stability analyses and observational constraints on SFDM oscillatons.
\end{abstract}

\maketitle

\section{Introduction}
\label{sec:1}

Dark matter (DM) a non-baryonic component of the Universe is invoked to explain a wide range of phenomena not accounted for by ordinary (baryonic) matter alone. Its presence across multiple length scales, from individual galaxies to galaxy clusters and up to the cosmological horizon (see, e.g., \cite{1}), has been a central subject in particle physics and modern cosmology for decades.

A wealth of dynamical and observational evidence strongly supports the existence of DM, including galaxy rotation curves (GRCs) \citep{2,3}, the mass content of galaxy clusters \cite{4,5}, gravitational lensing \citep{6,7}, distance redshift relations \cite{8}, bulk flows \cite{9}, large-scale structure (LSS) \citep{10}, and structure formation \citep{11,12}. Despite this evidence, the \emph{nature} and \emph{microphysical composition} of DM remain unknown. Consequently, guided by observations and theory, a broad spectrum of models has been developed, which can be grouped as follows:
\begin{itemize}
  \item \emph{Standard Model of Cosmology} (SMC; the concordance model): the $\Lambda$ cold dark matter ($\Lambda$CDM) framework \citep{13,14,15}, encompassing cold dark matter (CDM; often modeled as a pressureless fluid of weakly interacting massive particles, WIMPs), warm dark matter (WDM) \citep{16,17,18}, self-interacting dark matter \cite{19,20}, and related variants.
  \item \emph{Standard Model (SM) of particle physics} and its extensions: quantum-field-theoretic candidates \citep{21,22,23,24,25}, including models based on scalar fields (SFs) \citep{26,27,28,29,30}.
\end{itemize}
Within these approaches, proposed candidates span WIMPs to ultra-light scalars; see, e.g., \citep{24,25,31} and references therein.

At present, $\Lambda$CDM where CDM denotes cold dark matter and $\Lambda$ the cosmological constant serves as the leading cosmological paradigm. Rooted in inflationary initial conditions, general relativity, and the SM of particle physics, it successfully accounts for:
\begin{enumerate}
  \item the matter power spectrum and, in particular, the spectrum of temperature/polarization anisotropies in the cosmic microwave background (CMB);
  \item hierarchical structure formation via gravitational instability from primordial density perturbations;
  \item late-time accelerated expansion attributed to the cosmological constant $\Lambda$;
  \item the distribution of large-scale structures shaped by a weakly interacting CDM component.
\end{enumerate}
Nevertheless, open questions persist. For example, while WIMPs are among the simplest and most widely studied CDM candidates, non-gravitational interactions with SM particles have not been directly detected. Such considerations motivate alternative scenarios inspired by high-energy theory, including ultra-light scalar fields and axion-like particles (for overviews, see \citep{28,32,33,34,35}).

Among scalar-field (SF) scenarios, the scalar-field dark matter (SFDM) hypothesis models DM as a relativistic scalar field $\Phi$ with a suitable self-interaction potential $V(\Phi)$ \cite{36}. SFDM can reproduce many of the cosmological-scale predictions of $\Lambda$CDM while yielding distinctive phenomenology on galactic and sub-galactic scales. It also suggests that galaxy halos share common structural features and can form at relatively high redshifts compared with typical expectations in $\Lambda$CDM; if the SF constituents are ultra-light, SFDM aligns with the cases discussed in \cite{30}. SFDM models can be categorized according to whether the underlying scalar is complex or real. Complex scalar fields (CSFs) admit a global $U(1)$ symmetry and form gravitationally bound, stationary configurations \emph{boson stars} which have been extensively studied as astrophysical objects (see, e.g., \citep{37,38,39,40}). By contrast, real scalar fields (RSFs) can form \emph{oscillatons}: localized, time-periodic, self-gravitating configurations minimally coupled to gravity through a potential $V(\Phi)$ and interacting with other matter only gravitationally \cite{41,42}. Oscillatons are fully relativistic, bounded, and inherently dynamical, in contrast to the effectively stationary nature of many boson-star solutions. From the perspective of general relativity, they provide spherically symmetric, non-topological, non-singular, and asymptotically flat solutions of the coupled Einstein-Klein-Gordon (EKG) system \cite{43}.

In this work we investigate oscillatons within SFDM models featuring an \emph{exponential} self-interaction potential. Our formulation develops a full Fourier (Jacobi-Anger) expansion that resums the time dependence of the metric and the potential, thereby capturing quadratic, quartic, and higher-order contributions within a single framework. We present the mathematical setup for constructing time-periodic, spherically symmetric solutions; state the boundary conditions implementing regularity and asymptotic flatness; and describe the numerical approach and resulting profiles, including energy density, radial energy flux (rather than “momentum density”), pressure components, and mass, across a range of free parameters and boundary conditions.

Section~\ref{sec:2} provides the mathematical background on oscillatons required for our construction. Section~\ref{sec:3} introduces the specific potential considered here (full Fourier expansion of an exponential scalar potential), compiles useful analytical relations, and presents the boundary conditions along with numerical results. Conclusions are summarized in Section~\ref{sec:4}.

\section{Background of Oscillaton Mathematics}
\label{sec:2}

Following the seminal formulation in \cite{41}, we briefly review the essential equations governing spherically symmetric, time-dependent real scalar fields minimally coupled to gravity.

The most general spherically symmetric line element compatible with a time-dependent real scalar field (RSF) can be written as
\begin{equation}
  ds^2 \equiv g_{\alpha\beta}\,dx^{\alpha}dx^{\beta}
  = -e^{\nu(t,r)-\mu(t,r)}\,dt^2
    + e^{\nu(t,r)+\mu(t,r)}\,dr^2
    + r^2\!\left(d\theta^2+\sin^2\!\theta\,d\varphi^2\right),
\label{equ1}
\end{equation}
where $\nu=\nu(t,r)$ and $\mu=\mu(t,r)$ are metric functions. We work with signature $(-,+,+,+)$ and set $c=\hbar=1$.

For a real scalar field $\Phi(t,r)$ with potential $V(\Phi)$, the stress energy tensor is
\begin{equation}
  T_{\alpha\beta}
  = \Phi_{,\alpha}\Phi_{,\beta}
    - \frac{1}{2}\,g_{\alpha\beta}\!\left(\Phi^{,\gamma}\Phi_{,\gamma}+2V(\Phi)\right).
  \label{equ2}
\end{equation}
Using the metric \eqref{equ1}, the non-vanishing mixed components relevant to a $3{+}1$ decomposition are
\begin{align}
  -T^{0}{}_{0} \;\equiv\; \rho_{\Phi}
  &= \frac{1}{2}\!\left[e^{-(\nu-\mu)}\,\dot{\Phi}^{2} + e^{-(\nu+\mu)}\,\Phi'^{2} + 2V(\Phi)\right], \label{equ3}\\[0.3em]
  T^{1}{}_{0} \;\equiv\; j_{r}
  &= e^{-(\nu+\mu)}\,\dot{\Phi}\,\Phi', \label{equ4}\\[0.3em]
  T^{1}{}_{1} \;\equiv\; p_{r}
  &= \frac{1}{2}\!\left[e^{-(\nu-\mu)}\,\dot{\Phi}^{2} + e^{-(\nu+\mu)}\,\Phi'^{2} - 2V(\Phi)\right], \label{equ5}\\[0.3em]
  T^{2}{}_{2} = T^{3}{}_{3} \;\equiv\; p_{\perp}
  &= \frac{1}{2}\!\left[e^{-(\nu-\mu)}\,\dot{\Phi}^{2} - e^{-(\nu+\mu)}\,\Phi'^{2} - 2V(\Phi)\right], \label{equ6}
\end{align}
where an overdot denotes $\partial/\partial t$ and a prime denotes $\partial/\partial r$. Here $\rho_{\Phi}$ is the energy density, $j_r\,(=T^{r}{}_{0})$ is the radial \emph{energy flux} (often called “momentum density” in the literature), and $p_r$, $p_{\perp}$ are the radial and tangential pressures.

Einstein’s field equations,
\begin{equation}
  G_{\alpha\beta} \equiv R_{\alpha\beta} - \tfrac{1}{2}g_{\alpha\beta}R \;=\; k_{0}\,T_{\alpha\beta}, 
  \qquad k_{0}\equiv 8\pi G = \frac{8\pi}{M_{\mathrm{Pl}}^{2}},
\end{equation}
yield, for the metric \eqref{equ1}, the following independent relations for $\nu$ and $\mu$:
\begin{align}
  (\nu+\mu)^{\cdot} \;&=\; k_{0}\,r\,\dot{\Phi}\,\Phi', \label{equ7}\\
  \nu'(t,r) \;&=\; \frac{k_{0}\,r}{2}\!\left(e^{2\mu}\,\dot{\Phi}^{2} + \Phi'^{2}\right), \label{equ8}\\
  \mu'(t,r) \;&=\; \frac{1}{r}\!\left[\,1 + e^{\nu+\mu}\!\left(k_{0}\,r^{2}V(\Phi)-1\right)\right]. \label{equ9}
\end{align}
Equation \eqref{equ7} encodes the mixed $(t,r)$ component (a momentum/flux constraint), while \eqref{equ8} and \eqref{equ9} are the radial constraint/evolution equations consistent with spherical symmetry.

Stress energy conservation $\nabla_{\alpha}T^{\alpha}{}_{\beta}=0$ is equivalent to the covariant Klein-Gordon equation
\begin{equation}
  \Box\Phi - \frac{dV}{d\Phi} = 0,
  \qquad 
  \Box\Phi \equiv g^{\alpha\beta}\nabla_{\alpha}\nabla_{\beta}\Phi.
  \label{equ10}
\end{equation}
For the metric \eqref{equ1}, the d'Alembertian evaluates to
\begin{equation}
  \Box\Phi
  = -\,e^{-(\nu-\mu)}\!\left[\ddot{\Phi}
      + \tfrac{1}{2}\big(\dot{\nu}-\dot{\mu}\big)\dot{\Phi}\right]
    + e^{-(\nu+\mu)}\!\left[\Phi'' 
      + \left(\frac{2}{r} + \tfrac{1}{2}\big(\nu'+\mu'\big)\right)\Phi'\right],
  \label{equ11}
\end{equation}
so that \eqref{equ10} becomes the explicit PDE
\begin{equation}
  -\,e^{-(\nu-\mu)}\!\left[\ddot{\Phi}
      + \tfrac{1}{2}\big(\dot{\nu}-\dot{\mu}\big)\dot{\Phi}\right]
  + e^{-(\nu+\mu)}\!\left[\Phi'' 
      + \left(\frac{2}{r} + \tfrac{1}{2}\big(\nu'+\mu'\big)\right)\Phi'\right]
  - \frac{dV}{d\Phi} \;=\; 0.
  \label{equ12}
\end{equation}

Equations \eqref{equ7}--\eqref{equ12} hold for any potential $V(\Phi)$. Two commonly used cases are:
\begin{itemize}
  \item \textbf{Quadratic (free massive) field:} $V(\Phi)=\tfrac{1}{2}m_{\Phi}^{2}\Phi^{2}$, which underpins many small-amplitude oscillaton analyses and serves as a useful validation limit for numerical schemes.
  \item \textbf{Exponential self-interaction:} $V(\Phi)=V_{0}\,e^{-\lambda\sqrt{k_{0}}\,\Phi}$, with $V_{0}$ and $\lambda$ treated as free parameters constrained by observations and by consistency of solutions. No \emph{a priori} small- or large-amplitude assumption on $\lambda\sqrt{k_{0}}\Phi$ is required at the level of the covariant system; the regime of validity of any subsequent Fourier/Bessel expansion will be stated explicitly where used.
\end{itemize}

Within SFDM scenarios, a cosmological background of scalar quanta can seed localized, self-gravitating fluctuations which, depending on $V(\Phi)$ and initial data, relax into long-lived, time-periodic configurations \emph{oscillatons}. These objects provide halo-like cores or compact configurations in structure-formation settings (see, e.g., \citep{44,45} for related scenarios). In the remainder of this work we employ the above system to construct and analyze oscillaton solutions for exponential potentials via a controlled Fourier expansion, together with appropriate boundary conditions and numerical diagnostics.

\section{Complete Expansion of the Exponential Potential}
\label{sec:3}

In Section~\ref{sec:2} we summarized the covariant system governing a time-dependent real scalar field (RSF) in spherical symmetry. We now specialize to \emph{time-periodic} configurations and develop a controlled Fourier treatment for both the metric and the potential. 

\subsection{Ansatz, frequency, and dimensionless variables}
We assume a single fundamental angular frequency $\omega$ and write the scalar field as \cite{43}
\begin{equation}
  \Phi(t,r) \;=\; \frac{2\,\sigma(r)}{\sqrt{k_{0}}}\,\cos(\omega t),
  \label{eq:phi-ansatz}
\end{equation}
where $\sigma(r)$ is the radial amplitude. Introducing the dimensionless radius and frequency
\begin{equation}
  x \equiv m_{\Phi} r,
  \qquad
  \Omega \equiv \frac{\omega}{m_{\Phi}},
\end{equation}
Where $m_{\Phi}$ is considered as the mass of scalar particles. We henceforth regard all radial functions as functions of $x$, and use a prime ${}'$ to denote $d/dx$. With this convention the ansatz reads
\[
  \Phi(t,x)=\frac{2\,\sigma(x)}{\sqrt{k_{0}}}\cos(\omega t).
\]

Substituting \eqref{eq:phi-ansatz} into the momentum constraint \eqref{equ7} and integrating in time shows that $(\nu+\mu)$ contains only the even harmonic $2\omega$. Writing
\begin{equation}
  \nu(t,x) = \nu_{0}(x)+\nu_{1}(x)\cos(2\omega t),
  \qquad
  \mu(t,x) = \mu_{0}(x)+\mu_{1}(x)\cos(2\omega t),
  \label{eq:harm-metric}
\end{equation}
one finds the relation
\begin{equation}
  \nu_{1}(x)+\mu_{1}(x)=x\,\sigma(x)\,\sigma'(x),
  \label{eq:nuplusmu-harm}
\end{equation}
consistent with $\dot{\Phi}\,\Phi'\propto \sin(2\omega t)$ and $(\nu+\mu)^{\cdot}\propto \sin(2\omega t)$.

\subsection{Jacobi--Anger (modified Bessel) expansions}
For any function of the form $A(x)+B(x)\cos(2\omega t)$ we will repeatedly use
\begin{equation}
  e^{A+B\cos(2\omega t)} \;=\; e^{A}\Big[I_{0}(B) \;+\;2\sum_{n=1}^{\infty} I_{n}(B)\cos(2n\omega t)\Big],
  \label{eq:JA}
\end{equation}
where $I_{n}$ are modified Bessel functions of the first kind. Applying \eqref{eq:JA} to $e^{\nu\pm \mu}$ with \eqref{eq:harm-metric} yields
\begin{equation}
  e^{\nu\pm\mu}
  \;=\;
  e^{\nu_{0}\pm \mu_{0}}
  \Big[ I_{0}(\nu_{1}\pm\mu_{1}) + 2\sum_{n=1}^{\infty} I_{n}(\nu_{1}\pm\mu_{1})\cos(2n\omega t) \Big].
  \label{equ14}
\end{equation}
Likewise, for the \emph{exponential} potential
\begin{equation}
  V(\Phi) \;=\; V_{0}\,e^{-\lambda\sqrt{k_{0}}\,\Phi}
             \;=\; V_{0}\,\exp\!\big[-2\lambda\,\sigma(x)\cos(\omega t)\big],
  \label{eq:Vexp-section3}
\end{equation}
we have the cosine–Fourier series
\begin{equation}
  V(\Phi)
  \;=\;
  V_{0}\Big[ I_{0}\!\big(2\lambda\sigma\big) + 2\sum_{n=1}^{\infty} (-1)^{n} I_{n}\!\big(2\lambda\sigma\big)\cos(n\omega t) \Big],
  \label{eq:Vexp-Fourier}
\end{equation}
where $I_{0}(-z)=I_{0}(z)$ and $I_{n}(-z)=(-1)^{n}I_{n}(z)$ applied to $e^{-2\lambda\sigma\cos(\omega t)}$.

Expanding \eqref{eq:Vexp-section3} for small $\Phi$ gives
$V(\Phi)=V_{0}\big(1 - \lambda\sqrt{k_{0}}\Phi + \tfrac{1}{2}\lambda^{2}k_{0}\Phi^{2} + \cdots\big)$,
hence the quadratic mass term is
\begin{equation}
  m_{\Phi}^{2} \;\equiv\; V''(\Phi)\big|_{\Phi=0} \;=\; \lambda^{2}\,k_{0}\,V_{0}
  \quad\Longrightarrow\quad
  V_{0} \;=\; \frac{m_{\Phi}^{2}}{\lambda^{2}\,k_{0}}.
  \label{eq:V0-consistency}
\end{equation}
This relation ensures dimensional consistency and the correct small-amplitude limit; it replaces ad-hoc choices such as $V_{0}=m_{\Phi}^{2}/k_{0}$ (which would implicitly fix $\lambda=1$).

\subsection{Harmonic matching and rescalings}
Inserting \eqref{eq:phi-ansatz}, \eqref{equ14}, and \eqref{eq:Vexp-Fourier} into the Einstein equations \eqref{equ7}--\eqref{equ9} and the Klein--Gordon equation from Sec.~\ref{sec:2}, and equating the coefficients of $\{\cos(0\cdot \omega t),\cos(2\omega t)\}$ on both sides, yields a closed, time-independent system for $\{\sigma,\nu_{0},\mu_{0},\nu_{1},\mu_{1}\}$ as functions of $x$. To make the equations dimensionless, we use $x=m_{\Phi}r$, so that
\[
  \nu'(t,r)=m_{\Phi}\,\nu'(t,x),\quad
  \mu'(t,r)=m_{\Phi}\,\mu'(t,x),\quad
  \sigma'(r)=m_{\Phi}\,\sigma'(x),\quad
  \sigma''(r)=m_{\Phi}^{2}\,\sigma''(x).
\]
We also define $\Omega=\omega/m_{\Phi}$. With these conventions, the matched equations take the form

\begin{align}
  \sigma'' \;=\;&
  -\,\sigma'\!\left(\frac{2}{x} - \mu_{0}' - \frac{1}{2}\mu_{1}'\right)
  \nonumber\\
  &\;-\; e^{2\mu_{0}}\,\sigma\Big[\,I_{0}\!\big(2\mu_{1}\big)
      - I_{1}\!\big(2\mu_{1}\big) - \mu_{1}\,I_{2}\!\big(2\mu_{1}\big)\Big]
  \nonumber\\
  &\;+\; \lambda\,e^{\nu_{0}+\mu_{0}}
      \sum_{n=0}^{\infty} I_{2n+1}\!\big(2\lambda\sigma\big)
      \Big[ I_{n}\!\big(\nu_{1}+\mu_{1}\big) + I_{n+1}\!\big(\nu_{1}+\mu_{1}\big) \Big].
  \label{equ14-clean}
\end{align}

\begin{align}
  \nu_{0}' \;=\;& x\Big\{ e^{2\mu_{0}}\,\sigma^{2}\big[\,I_{0}\!\big(2\mu_{1}\big)-I_{1}\!\big(2\mu_{1}\big)\big]
                    \;+\; (\sigma')^{2}\Big\},
  \label{equ15-clean}
  \\[0.25em]
  \nu_{1}' \;=\;& x\Big\{ e^{2\mu_{0}}\,\sigma^{2}\big[-I_{0}\!\big(2\mu_{1}\big)+2I_{1}\!\big(2\mu_{1}\big)-I_{2}\!\big(2\mu_{1}\big)\big]
                    \;+\; (\sigma')^{2}\Big\}.
  \label{equ16-clean}
\end{align}

\begin{align}
  \mu_{0}' \;=\;& \frac{1}{x}\Bigg\{1 + e^{\nu_{0}+\mu_{0}}
      \Bigg[
         x^{2}\!\left(I_{0}\!\big(\nu_{1}+\mu_{1}\big)\,I_{0}\!\big(2\lambda\sigma\big)
              + 2\sum_{n=1}^{\infty} I_{n}\!\big(\nu_{1}+\mu_{1}\big)\,I_{2n}\!\big(2\lambda\sigma\big)\right)
         - I_{0}\!\big(\nu_{1}+\mu_{1}\big)
      \Bigg]\Bigg\},
  \label{equ17-clean}
  \\[0.25em]
  \mu_{1}' \;=\;& \frac{2}{x}\,e^{\nu_{0}+\mu_{0}}
     \Bigg[
        x^{2}\!\left( I_{0}\!\big(\nu_{1}+\mu_{1}\big)\,I_{2}\!\big(2\lambda\sigma\big)
           + \sum_{n=0}^{\infty} I_{n+1}\!\big(\nu_{1}+\mu_{1}\big)
               \big(I_{2n}\!\big(2\lambda\sigma\big)+I_{2n+4}\!\big(2\lambda\sigma\big)\big)\right)
        - I_{1}\!\big(\nu_{1}+\mu_{1}\big)
     \Bigg].
  \label{equ18-clean}
\end{align}
Equations \eqref{equ14-clean}--\eqref{equ18-clean} constitute a closed system when supplemented with \eqref{eq:nuplusmu-harm}.
They are exact within the harmonic content of \eqref{eq:harm-metric} and the Bessel resummations \eqref{equ14}--\eqref{eq:Vexp-Fourier}. In practice, the infinite sums are truncated at some $N$ once numerical convergence in $N$ is demonstrated.

\begin{flushleft}
\textbf{Remarks.}
\end{flushleft}
\begin{itemize}
  \item The terms proportional to $\Omega^{2}$ arise from $\dot{\Phi}^{2}$ and carry factors of $e^{2\mu_{0}}$ and Bessel combinations in $\mu_{1}$, while the gradient terms $(\sigma')^{2}$ originate from $\Phi'^{2}$ and contribute to both the mean and $2\omega$ sectors.
  \item The Bessel arguments $\nu_{1}\pm \mu_{1}$ and $2\lambda\sigma$ reflect, respectively, the time dependence sourced by the metric exponent and by the exponential potential. Using \eqref{eq:nuplusmu-harm}, one can equivalently express all occurrences of $(\nu_{1}+\mu_{1})$ in terms of $x\,\sigma\sigma'$.
  \item The mass normalization \eqref{eq:V0-consistency} is essential for a consistent small-amplitude limit and removes an otherwise spurious degree of freedom.
\end{itemize}


\subsection{Boundary conditions}
\label{subsec3-2}

We impose regularity at the origin and asymptotic flatness at spatial infinity. With the time periodic ansatz
\[
\Phi(t,x)=\frac{2\,\sigma(x)}{\sqrt{k_{0}}}\cos(\omega t),\qquad
\nu(t,x)=\nu_{0}(x)+\nu_{1}(x)\cos(2\omega t),\qquad
\mu(t,x)=\mu_{0}(x)+\mu_{1}(x)\cos(2\omega t),
\]
and $x\equiv m_\Phi r$, the requirements are as follows.

\paragraph{Regular center ($x\to 0$).}
\begin{equation}
\sigma'(0)=0,\qquad \mu_{0}(0)=-\nu_{0}(0),\qquad
\nu_{1}(0)=-\mu_{1}(0)
\label{eq:BC-center-1}
\end{equation}
with finite $\nu_{0}(0)$ and $\mu_{0}(0)$. Using the kinematic constraint
\begin{equation}
\nu_{1}(x)+\mu_{1}(x)=x\,\sigma(x)\,\sigma'(x),
\label{eq:constraint-sum-BC}
\end{equation}
we obtain $\nu_{1}+\mu_{1}=\mathcal{O}(x^{2})$ near the center, consistent with \eqref{eq:BC-center-1}. Without loss of generality (gauge choice) the relation, $\mu_{0}(0)=-\nu_{0}(0)$,
fixes the local orthonormal normalization at the center. In practice, regularity is enforced by a small $x$ Taylor expansion:
\[
\sigma(x)=\sigma_c+\tfrac{1}{2}\sigma_2 x^{2}+\mathcal{O}(x^{4}),\quad
\nu_{1}(x)=\tfrac{1}{2}\nu_{1,2}x^{2}+\mathcal{O}(x^{4}),\quad
\mu_{1}(x)=\tfrac{1}{2}\mu_{1,2}x^{2}+\mathcal{O}(x^{4}),
\]
with coefficients constrained by \eqref{equ14-clean}-\eqref{equ18-clean}.

\paragraph{Asymptotics ($x\to\infty$).}
\begin{equation}
\sigma(\infty)=0,\qquad
\nu_{1}(\infty)=0,\qquad
\mu_{1}(\infty)=0,
\label{eq:BC-infty-1}
\end{equation}
and time normalization chosen so that the metric approaches Minkowski,
\begin{equation}
\Omega^{2} e^{\nu_{0}(\infty)-\mu_{0}(\infty)}=1
\quad\Longrightarrow\quad
\mu_{0}(\infty)=-\nu_{0}(\infty)\neq 0,
\label{eq:BC-infty-2}
\end{equation}
i.e.\ $g_{tt}\to -1$ and $g_{rr}\to +1$ as $x\to\infty$; the physical frequency is then $\omega=\Omega\,m_\Phi$.

\paragraph{Free data and uniqueness.}
For given central amplitude $\sigma_c\equiv\sigma(0)$ and coupling $\lambda$, the system \eqref{equ14-clean}--\eqref{equ18-clean} with \eqref{eq:BC-center-1}--\eqref{eq:BC-infty-2} defines a boundary value problem. We shoot on the eigenfrequency $\Omega$ (and, if desired, one central metric value before enforcing \eqref{eq:BC-center-1}) to satisfy \eqref{eq:BC-infty-1}-\eqref{eq:BC-infty-2}. For fixed $(\sigma_c,\lambda)$ we find at most one $\Omega$ yielding an asymptotically flat, nodeless ground state; nodeful excited states also exist but are typically less stable.

\paragraph{Notes.}
(i) No \emph{a priori} requirement such as $|2\lambda\sigma|\gg 1$ is imposed by the covariant equations; the Fourier-Bessel truncation is controlled numerically by increasing the cutoff until observables converge. (ii) The frequently used relation $\nu_{0}(0)=-\mu_{0}(0)$ is a gauge choice; this choice is equivalent up to a constant shift and simplifies implementation.

\subsection{Numerical results and figure interpretations}
\label{subsec3-3}

We solve the coupled ordinary differential equations (ODEs) system \eqref{equ14-clean}--\eqref{equ18-clean} on $x\in[0,x_{\max}]$ with the boundary conditions in Sec.~\ref{subsec3-2}. The infinite Bessel sums are truncated at order $N$ and increased until convergence. Our pipeline is:

\begin{enumerate}
  \item \textbf{Discretization:} fourth  to sixth order finite differences on a nonuniform grid clustered near $x=0$. Choose $x_{\max}$ so that $\max\{|\nu_0|,|\mu_0|,|\nu_1|,|\mu_1|,|\sigma|\}<10^{-8}$ at $x_{\max}$.
  \item \textbf{Nonlinear solve:} shooting on $\Omega$ and Newton-Krylov relaxation for $\mathbf{u}=(\sigma,\nu_0,\mu_0,\nu_1,\mu_1)$, enforcing \eqref{eq:constraint-sum-BC} algebraically.
  \item \textbf{Convergence tests:} refine $(N\to N{+}2)$, $(\Delta x\to \Delta x/2)$, and increase $x_{\max}$ until global quantities (mass $M$, effective radius $R_{\rm eff}$) and local profiles change by $<10^{-4}$ relative.
\end{enumerate}

The metric prefactors follow from the Jacobi-Anger expansions,
\begin{align}
e^{\nu-\mu}
&= e^{\nu_{0}-\mu_{0}}
   \!\left[ I_{0}(\nu_{1}-\mu_{1})
          + 2\sum_{n=1}^{\infty} I_{n}(\nu_{1}-\mu_{1})\cos(2n\omega t) \right]
   \equiv g_{tt,0}(x) + g_{tt,2}(x)\cos(2\omega t) + g_{tt,4}(x)\cos(4\omega t)+\cdots, 
\label{equ20}
\\
e^{\nu+\mu}
&= e^{\nu_{0}+\mu_{0}}
   \!\left[ I_{0}(\nu_{1}+\mu_{1})
          + 2\sum_{n=1}^{\infty} I_{n}(\nu_{1}+\mu_{1})\cos(2n\omega t) \right]
   \equiv g_{rr,0}(x) + g_{rr,2}(x)\cos(2\omega t) + g_{rr,4}(x)\cos(4\omega t)+\cdots,
\label{equ21}
\end{align}
so the geometry carries only \emph{even} harmonics of $\omega$. By contrast, the scalar field oscillates at the fundamental $\omega$ and the exponential potential admits a full cosine series (Eq.~\eqref{eq:Vexp-Fourier}); when forming composite observables, this structure yields a dominant DC part plus even harmonics, while the radial flux peaks at $2\omega$.

\paragraph{Example setup.}
Unless stated otherwise we use $\sigma(0)=\frac{1.2}{\sqrt{2}}$, $\lambda=1$, and tune $\Omega$ to meet \eqref{eq:BC-infty-1}-\eqref{eq:BC-infty-2}. We fix the gauge via $\nu_0(0)=-\mu_0(0)=0.7855$; any alternative finite choice is equivalent up to a constant shift.

\begin{figure}[h]
\centering 
\includegraphics[width=0.45\textwidth]{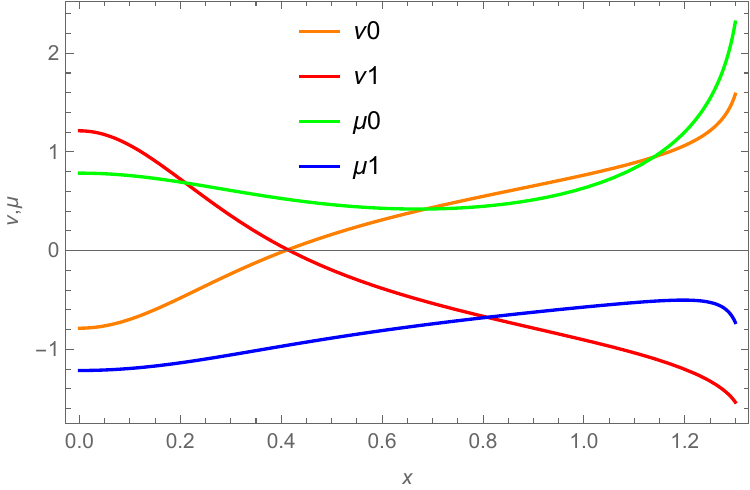}
\hfill
\includegraphics[width=0.45\textwidth]{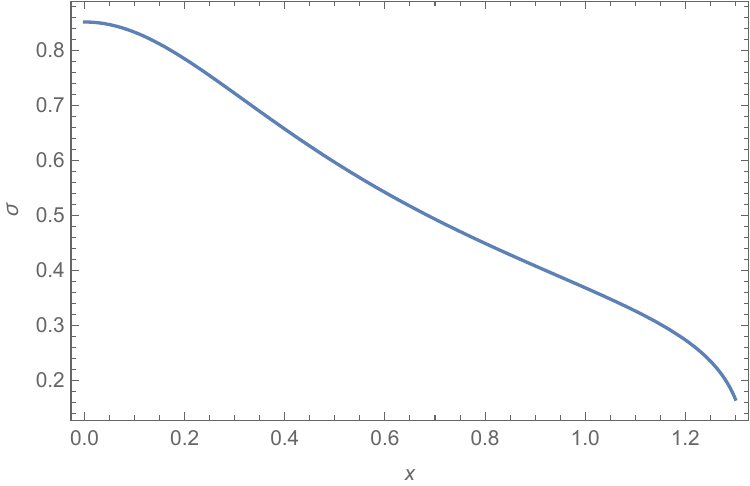}
\caption{Radial profiles for the time-periodic oscillaton with central amplitude $\sigma(0)=\frac{1.2}{\sqrt{2}}$ and coupling $\lambda=1$ (gauge: $\nu_0(0)=-\mu_0(0)=0.7855$; $\Omega$ tuned to enforce asymptotic flatness).
\textbf{Left:} time-averaged metric exponents $\nu_0(x),\mu_0(x)$ and leading even-harmonic amplitudes $\nu_1(x),\mu_1(x)$.
\textbf{Right:} scalar radial amplitude $\sigma(x)$.
Regularity imposes $\sigma'(0)=0$ and $\nu_1(0)=\mu_1(0)=1.2141$, while the constraint $\nu_1+\mu_1=x\,\sigma\sigma'$ is satisfied to solver tolerance.}
\label{fig:1}
\end{figure}

\paragraph{Figure~\ref{fig:1} (metrics functions or field vs.\ radius).}
\emph{Left:} metric functions, $\nu_0(x),\mu_0(x)$, and also leading even-harmonic, $\nu_1(x),\mu_1(x)$, components.
\emph{Right:} scalar radial amplitude $\sigma(x)$.
Regularity enforces $\sigma'(0)=0$ and $\nu_1(0)=-\mu_1(0)$; all profiles decay exponentially (or smoother) as $x\to\infty$. The constraint \eqref{eq:nuplusmu-harm} implies $\nu_1+\mu_1=x\,\sigma\sigma'$ and is satisfied to solver tolerance.

\begin{figure}[h]
\centering 
\includegraphics[width=0.45\textwidth]{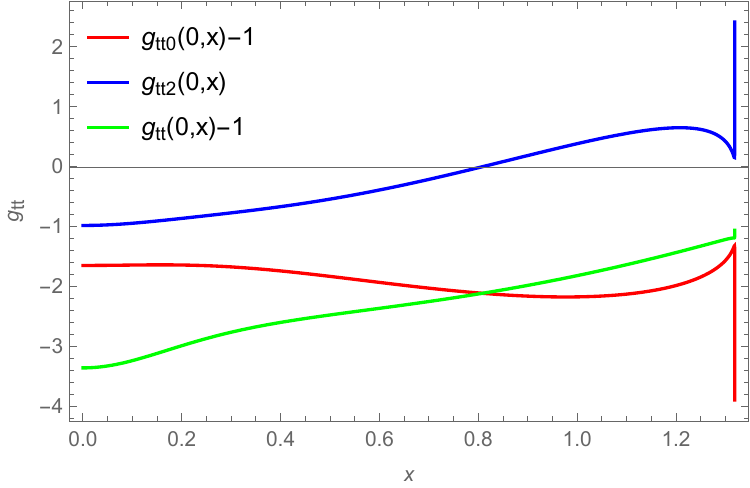}
\hfill
\includegraphics[width=0.45\textwidth]{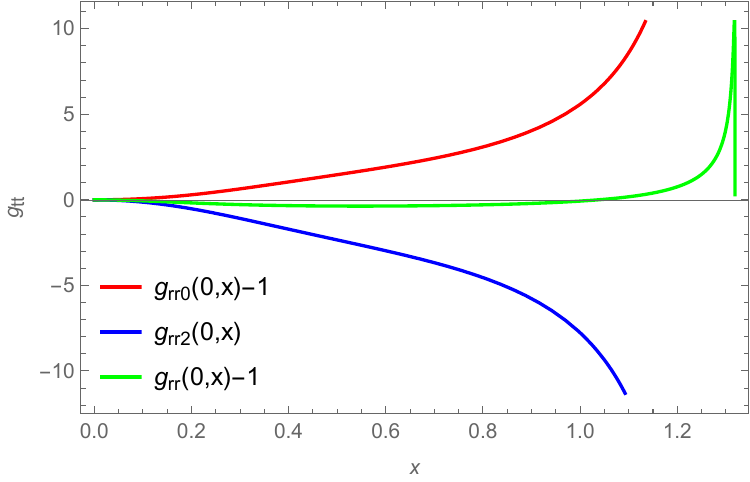}
\caption{Metric coefficients at $t=0$ and their leading harmonic content from Eqs.~\eqref{equ20}-\eqref{equ21} for $\sigma(0)=\frac{1.2}{\sqrt{2}}$, $\lambda=1$.
\textbf{Left:} $-g_{tt}(0,x)-1$ (green) together with $-g_{tt,0}(x)-1$ (blue) and $-g_{tt,2}(x)$ ($2\omega$–harmonic).
\textbf{Right:} $g_{rr}(0,x)-1$ (green) with $g_{rr,0}(x)-1$ and $g_{rr,2}(x)$ (blue).
The $2\omega$ mode dominates the time dependence; higher even harmonics are subleading once the Bessel truncation converges.}
\label{fig:2}
\end{figure}

\paragraph{Figure~\ref{fig:2} (metric coefficients and leading harmonics at $t=0$).}
\emph{Left:} $-g_{tt}(0,x)-1$ together with $-g_{tt,0}(x)-1$ and $-g_{tt,2}(x)$.
\emph{Right:} $g_{rr}(0,x)-1$ together with $g_{rr,0}(x)-1$ and $g_{rr,2}(x)$.
The $2\omega$ term dominates the time dependence; higher even harmonics ($4\omega$, $6\omega$, …) are subleading once the Bessel truncation is converged.

\begin{figure}[h]
\centering 
\includegraphics[width=0.45\textwidth]{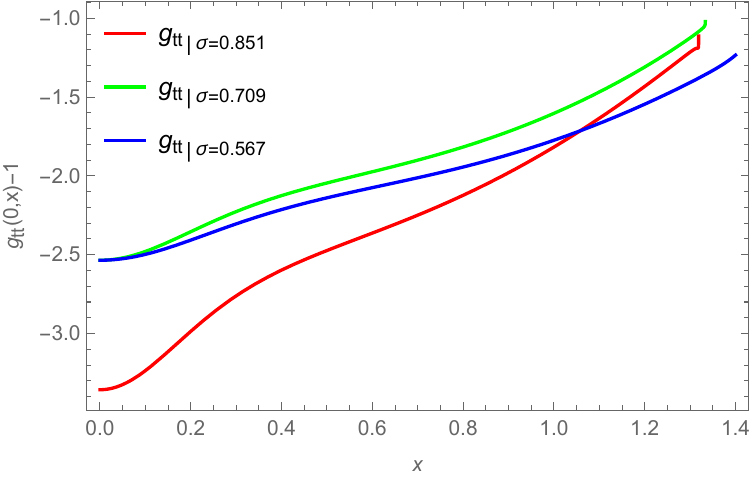}
\hfill
\includegraphics[width=0.45\textwidth]{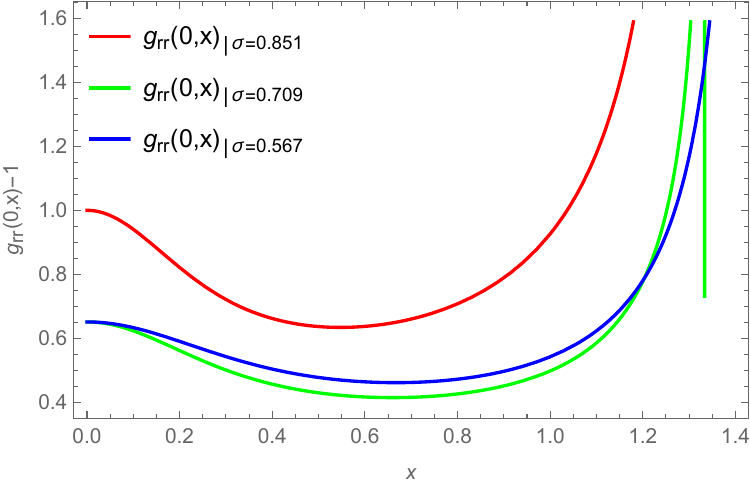}
\caption{Dependence of the geometry on the self-interaction strength $\sigma$ with other boundary data fixed.
We compare $-g_{tt}(0,x)-1$ (left) and $g_{rr}(0,x)-1$ (right) for $\sigma=\{0.851,709,567\}$.
Increasing $\sigma$ enhances $I_n(2\lambda\sigma)$, leading to larger departures from Minkowski and steeper core gradients, a trend robust under truncation/mesh refinement.}
\label{fig:3}
\end{figure}

\paragraph{Figure~\ref{fig:3} (dependence on the self-interaction $\sigma$).}
We compare $-g_{tt}(0,x)-1$ and $g_{rr}(0,x)-1$ for $\sigma\in\{0.8510,0.7092,0.5673\}$ with other boundary data fixed. Larger $\sigma$ enhances $I_n(2\lambda\sigma)$, increasing departures from Minkowski and steepening the core gradients. This trend is consistent with stronger self-interaction and survives truncation/mesh refinement checks.

\begin{figure}[t!]
\centering 
\includegraphics[width=0.45\textwidth]{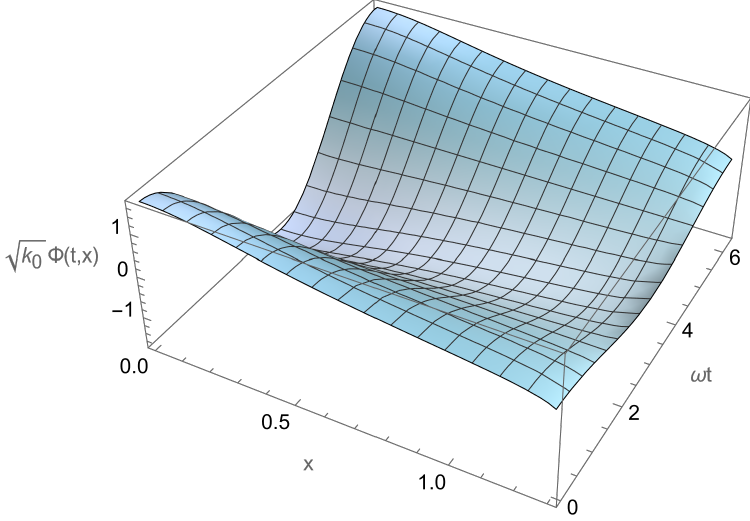}
\hfill
\includegraphics[width=0.45\textwidth]{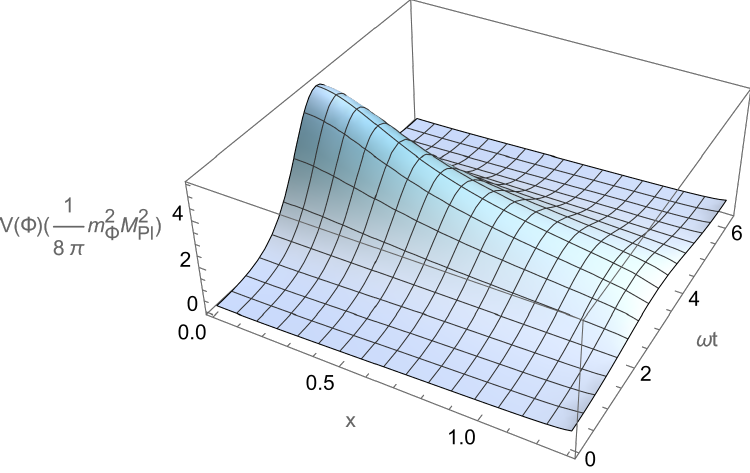}
\caption{Time dependence of the scalar and the exponential potential over one period for $\sigma(0)=\frac{1.2}{\sqrt{2}}$, $\lambda=1$ and $x\in(10^{-4},1.3]$.
\textbf{Left:} $\sqrt{k_0}\Phi(t,x)=2\,\sigma(x)\cos(\omega t)$ shows the fundamental $\omega$ oscillation modulated by the radial envelope.
\textbf{Right:} $V(\Phi)=V_0 e^{-\lambda\sqrt{k_0}\Phi}$ exhibits the full harmonic content implied by Eq.~\eqref{eq:Vexp-Fourier}; the small amplitude limit is consistent with $V_0=m_\Phi^2/(\lambda^2 k_0)$ from Eq.~\eqref{eq:V0-consistency}.}
\label{fig:4}
\end{figure}

\paragraph{Figure~\ref{fig:4} (time dependence of $\Phi$ and $V$).}
\emph{Left:} 3D surface of $\sqrt{k_0}\Phi(t,x)=2\sigma(x)\cos(\omega t)$ over $x\in [10^{-4},1.3]$ and one period ($\omega t \in [0,2\pi]$). The nodal structure in time follows the fundamental $\omega$.
\emph{Right:} 3D surface of $V(\Phi)=V_0 e^{-\lambda\sqrt{k_0}\Phi}$ showing its full harmonic content per Eq.~\eqref{eq:Vexp-Fourier} (including the alternating $(-1)^n$ factor). The small amplitude limit is governed by \eqref{eq:V0-consistency}, ensuring the correct quadratic mass $m_\Phi$.

\begin{figure}[]
\centering 
\includegraphics[width=0.45\textwidth]{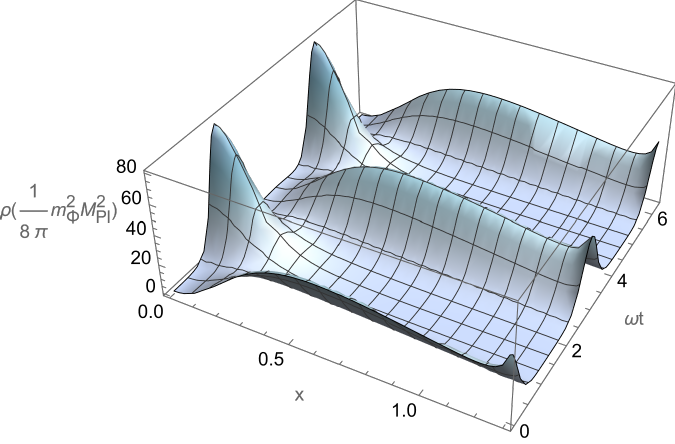}
\hfill
\includegraphics[width=0.45\textwidth]{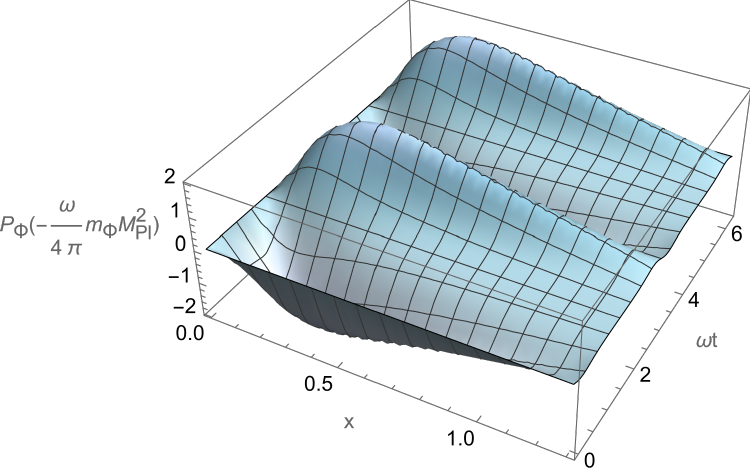}
\caption{Energy density and radial energy flux  for $\sigma(0)=\frac{1.2}{\sqrt{2}}$, $\lambda=1$.
\textbf{Left:} $\rho_\Phi(t,x)=\tfrac12\!\big[e^{-(\nu-\mu)}\dot{\Phi}^{2}+e^{-(\nu+\mu)}\Phi'^{2}+2V(\Phi)\big]$ has a dominant DC part plus even harmonics, peaking near the core and decaying rapidly with $x$.
\textbf{Right:} $j_r(t,x)=e^{-(\nu+\mu)}\dot{\Phi}\,\Phi'$ oscillates predominantly at $2\omega$ with the expected phase shift; it averages to zero over one period at fixed $x$, indicating no net mass transport in the stationary oscillaton.}
\label{fig:5}
\end{figure}

\paragraph{Figure~\ref{fig:5} (energy density and radial energy flux).}
We plot the \emph{energy density}
\[
\rho_{\Phi}=\tfrac{1}{2}\!\left[e^{-(\nu-\mu)}\dot{\Phi}^{2}+e^{-(\nu+\mu)}\Phi'^{2}+2V(\Phi)\right]
\]
(\emph{left}) and the \emph{radial energy flux}
\[
j_{r}=\,e^{-(\nu+\mu)}\,\dot{\Phi}\,\Phi'
\]
(\emph{right}). The density exhibits a DC component plus even harmonics, peaking near the core and decaying rapidly with $x$. The flux oscillates predominantly at $2\omega$ with the expected phase shift relative to $\Phi$; it integrates to zero over one period at fixed $x$ (no net mass transport in the time average for the stationary oscillaton).

\begin{figure}[h!]
\centering 
\includegraphics[width=0.45\textwidth]{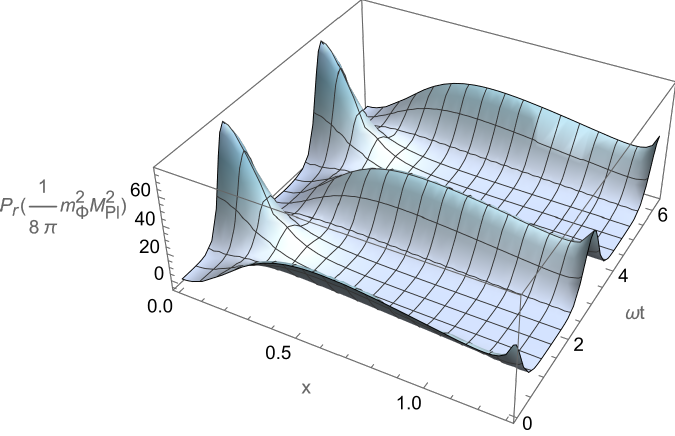}
\hfill
\includegraphics[width=0.45\textwidth]{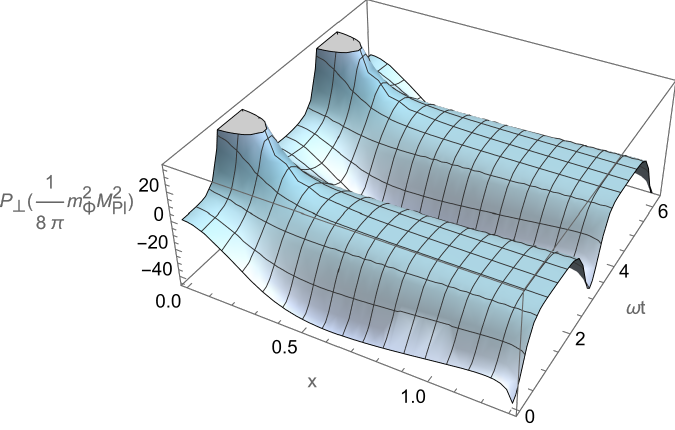}
\caption{Radial and tangential pressures  for $\sigma(0)=\frac{1.2}{\sqrt{2}}$, $\lambda=1$.
\textbf{Left:} $p_r(t,x)=\tfrac12\!\big[e^{-(\nu-\mu)}\dot{\Phi}^{2}+e^{-(\nu+\mu)}\Phi'^{2}-2V(\Phi)\big]$.
\textbf{Right:} $p_\perp(t,x)=\tfrac12\!\big[e^{-(\nu-\mu)}\dot{\Phi}^{2}-e^{-(\nu+\mu)}\Phi'^{2}-2V(\Phi)\big]$.
Instantaneous negative values can appear over parts of the cycle due to coherent phase relations among the even metric harmonics and the $\omega$ harmonic of $\Phi$; their GR-consistent interpretation is discussed in Sec.~\ref{subsec:energy-conditions} via WEC/NEC/SEC diagnostics.}
\label{fig:6}
\end{figure}

\paragraph{Figure~\ref{fig:6} (radial and tangential pressures).}
We display
\[
p_{r}=\tfrac{1}{2}\!\left[e^{-(\nu-\mu)}\dot{\Phi}^{2}+e^{-(\nu+\mu)}\Phi'^{2}-2V(\Phi)\right],
\qquad
p_{\perp}=\tfrac{1}{2}\!\left[e^{-(\nu-\mu)}\dot{\Phi}^{2}-e^{-(\nu+\mu)}\Phi'^{2}-2V(\Phi)\right]
\] Instantaneous negative values can occur over parts of the cycle due to coherent phase relations among the even metric harmonics and the $\omega$ harmonic of $\Phi$; their interpretation is given in Sec.~\ref{subsec:energy-conditions} via WEC/NEC/SEC diagnostics (no exotic matter is implied).

\paragraph{Averaging, truncation control, and global quantities.}
Using \eqref{equ20}-\eqref{equ21} and the Fourier series of $V(\Phi)$, we express $\rho_{\Phi},j_{r},p_{r},p_{\perp}$ as cosine series in $n\omega t$ and compute period averages
$\langle \rho_{\Phi}\rangle_{t}$, $\langle p_{r}\rangle_{t}$, $\langle p_{\perp}\rangle_{t}$.
The exponential potential is treated \emph{non-perturbatively} via Bessel resummations; there is no formal requirement like $|2\lambda\sigma|\gg1$. Numerically, we choose $N$ such that the tail
\[
\mathcal{R}_{N}(x)\equiv 2\sum_{n=N+1}^{\infty} I_{n}\!\big(2\lambda\sigma(x)\big)
\]
is uniformly $<10^{-8}$. The total mass $M$ follows from $g_{tt}$ asymptotics or from $\langle\rho_\Phi\rangle_t$. An effective radius $R_{\rm eff}$ is defined from a cumulative mass fraction (e.g., $90\%$). Stability proxies include the turning-point criterion $dM/d\sigma_c$ and long-time evolutions (beyond the scope of this static construction).

\subsection{Energy conditions and the interpretation of negative pressures}
\label{subsec:energy-conditions}

We evaluate the classical energy conditions for the oscillaton solutions constructed above (see, e.g., \cite{46,47,48,49}). Instantaneous observables namely the energy density $\rho_\Phi$, the radial energy flux $j_r$, and the radial/tangential pressures $p_r$ and $p_\perp$ are evaluated directly from the stress energy tensor constructed out of $\Phi$, $\dot{\Phi}$, $\Phi'$, the metric prefactors $e^{\nu\pm\mu}$, and the potential $V(\Phi)$; by contrast, their period-averaged counterparts  are defined over the fundamental period of the stress tensor, $T_s=\pi/\omega$, via
\begin{equation}
\langle X\rangle_t \;\equiv\; \frac{\omega}{\pi}\int_{0}^{\pi/\omega}\! X(t,x)\,dt.
\end{equation}
Then the components of stress energy tensor up to any desired order of expansion, in this text up to order four/six for $e^{-(\nu\pm\mu)}$/$e^{-2\lambda\sigma}$, respectively can be obtained as:
 \begin{align}
 \nonumber
 \langle\rho_{\Phi(t,x)}\rangle_t\;=&\dfrac{m_{\Phi}^2 M_{Pl}^2}{8\pi}\Bigg\lbrace\Big(\sigma^2 e^{-(\nu_0-\mu_0)}\big[I_0+I_1]_{(\nu_1-\mu_1)}+\sigma'^2\,e^{-(\nu_0+\mu_0)}\big[I_0- I_1\big]_{(\nu_1+\mu_1)}+I_0(2\sigma)\Big) \\ 
 \nonumber
 &-\dfrac{1}{\pi}\int_0^{\frac{2\pi}{\omega}}I_1(2\sigma)\cos(\omega t)\, dt \\
 \nonumber
&+\dfrac{1}{2\pi} \int_0^{\frac{\pi}{\omega}}\Big(-\sigma^{2}e^{-(\nu_0-\mu_0)}\big[I_0+2I_1+I_2\big]_{(\nu_1-\mu_1)}+\sigma'^{2}e^{-(\nu_0+\mu_0)}\big[I_0-2I_1+I_2\big]_{(\nu_1+\mu_1)}+2I_2(2\sigma)\Big)\cos(2\omega t)\,dt \\ 
\nonumber
&-\dfrac{1}{\pi}\int_0^{\frac{2\pi}{3\omega}}I_{3}(2\sigma)\cos(3\omega t)\, dt \\ 
\nonumber
&+\dfrac{1}{2\pi} \int_{0}^{\frac{\pi}{2\omega}}\Big(-\sigma^{2}e^{-(\nu_0-\mu_0)}\big[I_1+2I_2+I_3\big]_{(\nu_1-\mu_1)}-\sigma'^{2}e^{-(\nu_0+\mu_0)}\big[I_1-2I_2+I_3\big]_{(\nu_1+\mu_1)}+2I_4(2\sigma)\Big)\cos(4\omega t)\,dt \\
 \nonumber
&-\dfrac{1}{\pi}\int_0^{\frac{2\pi}{5\omega}}I_{5}(2\sigma)\cos(5\omega t)\, dt\\
\nonumber
&+\dfrac{1}{2\pi} \int_{0}^{\frac{\pi}{3\omega}}\Big(-\sigma^{2}e^{-(\nu_0-\mu_0)}\big[I_2+2I_3+I_4\big]_{(\nu_1-\mu_1)}+\sigma'^{2}e^{-(\nu_0+\mu_0)}\big[I_2-2I_3+I_4\big]_{(\nu_1+\mu_1)}+2I_6(2\sigma)\Big)\cos(6\omega t)\,dt \Bigg\rbrace\,\\
=&\dfrac{m_{\Phi}^{2}M_{Pl}^{2}}{8\pi}\Bigg\lbrace\sigma^{2}e^{-(\nu_0-\mu_0)}\big[I_0+I_1]_{(\nu_1-\mu_1)}+\sigma'^2e^{-(\nu_0+\mu_0)}\big[I_0- I_1\big]_{(\nu_1+\mu_1)}+I_0(2\sigma)\Bigg\rbrace,
\label{equ35}
 \end{align}
 \begin{align}
 \nonumber
\langle j_r(t,x)\rangle_t=-\dfrac{\omega\,m_\Phi\,M_{Pl}^{2}}{4\pi}\Bigg\lbrace \dfrac{1}{\pi}\int_0^{\frac{\pi}{\omega}}e^{-(\nu_0+\mu_0)}\big[I_0-I_2\big]_{(\nu_0+\mu_0)}\sin(2\omega t) \,dt
 -\dfrac{1}{\pi}\int_0^{\frac{\pi}{2\omega}} e^{-(\nu_0+\mu_0)}\big[I_1-I_3\big]_{(\nu_0+\mu_0)}\sin(4\omega t)\,dt\\
   +\dfrac{1}{\pi}\int_0^{\frac{\pi}{3\omega}}e^{-(\nu_{0}+\mu_{0})}\big[I_2-I_4\big]_{(\nu_0+\mu_0)}\sin(6\omega t)\,dt
 - \dfrac{1}{\pi}\int_0^{\frac{\pi}{4\omega}}-e^{-(\nu_{0}+\mu_{0})}\big[I_3-I_5\big]_{(\nu_0+\mu_0)}\sin(8\omega t)\,dt
 \Bigg\rbrace=0,
 \label{equ36}
 \end{align}
finally, radial/tangential pressure period-averaged relations, similar to Eq. \eqref{equ35},  leads to the following graphs: 

\begin{figure}[h]
\centering 
\includegraphics[width=0.45\textwidth]{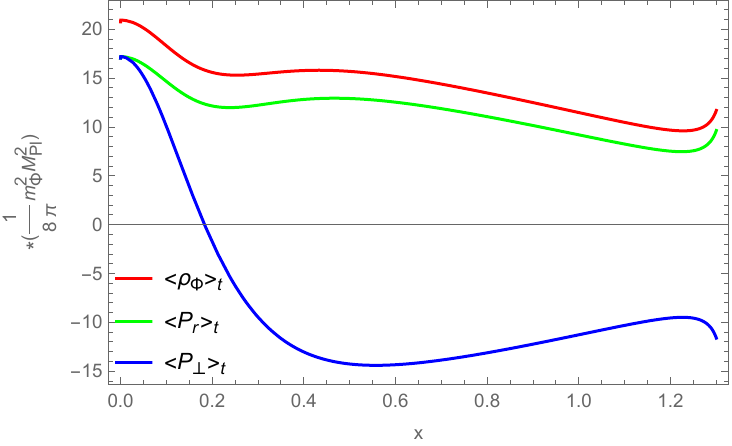}
\hfill
\includegraphics[width=0.45\textwidth]{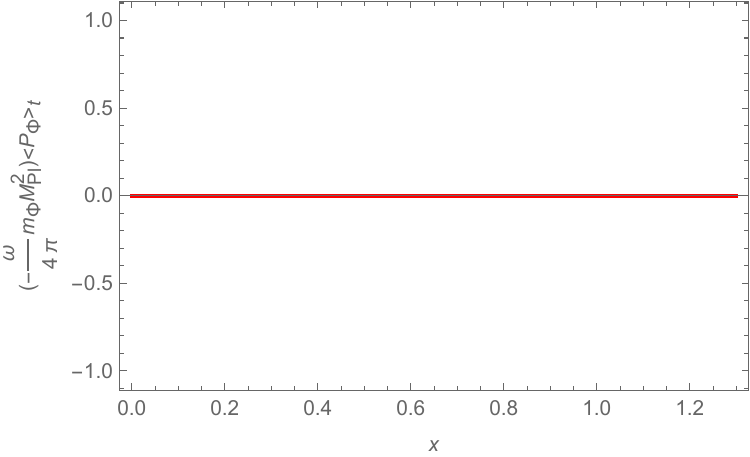}
\caption{Period-averaged counterparts of energy densiity/flux and Radia/tangential pressures  for $\sigma(0)=\frac{1.2}{\sqrt{2}}$, $\lambda=1$.
\textbf{Left:} $\langle\,\rho_r(t,x)\rangle_t=\langle\tfrac12\!\big[e^{-(\nu-\mu)}\dot{\Phi}^{2}+e^{-(\nu+\mu)}\Phi'^{2}+2V(\Phi)\big]\rangle_t$,$\langle\,P_r(t,x)\rangle_t=\langle\tfrac12\!\big[e^{-(\nu-\mu)}\dot{\Phi}^{2}+e^{-(\nu+\mu)}\Phi'^{2}-2V(\Phi)\big]\rangle_t$ and $\langle\,P_\perp(t,x)\rangle_t=\langle\tfrac12\!\big[e^{-(\nu-\mu)}\dot{\Phi}^{2}-e^{-(\nu+\mu)}\Phi'^{2}-2V(\Phi)\big]\rangle_t$.
\textbf{Right:} $\langle j_r(t,x)\rangle_t=\langle\, e^{-(\nu+\mu)}\dot{\Phi}\Phi'\rangle_t$.
 The GR interpretation of negative parts of quantities can be discussed in Sec.~\ref{subsec:energy-conditions} via WEC/NEC/SEC diagnostics. The average energy flux for dominate even frequencies between $2\omega\,t$ and $8\omega\,t$, is zero at fixed x, indicating that there is no net mass transfer in a stationary oscillaton.} 
\label{fig:7}
\end{figure}

In a local orthonormal tetrad adapted to $(t,r,\theta,\varphi)$ we use the standard pointwise inequalities (see for example \citep{50} and references therein)
\begin{align*}
 \text{WEC:}\;& \rho_\Phi \ge 0,\quad \rho_\Phi + p_r \ge 0,\quad \rho_\Phi + p_\perp \ge 0,\\
 \text{NEC:}\;& \rho_\Phi + p_r \ge 0,\quad \rho_\Phi + p_\perp \ge 0,\\
 \text{SEC:}\;& \rho_\Phi + p_r + 2p_\perp \ge 0\ \text{and NEC},\\
\text{DEC:}\;& \rho_\Phi \ge |p_r|,\quad \rho_\Phi \ge |p_\perp|.
\end{align*}
 When $j_r\neq0$ instantaneously, we evaluate the conditions in the local rest (Landau) frame-i.e., the instantaneous orthonormal frame in which $T^{\hat{\alpha}}{}_{\hat{\beta}}$ is diagonal (obtained by a radial boost that eliminates the flux) so that the above inequalities apply to the principal energy density and pressures.

For the time periodic solutions obtained via the Fourier-Bessel construction, we find:
(i) $\rho_\Phi(t,x)\!\ge\!0$ pointwise in our numerical solutions;
(ii) $p_r(t,x)$ and/or $p_\perp(t,x)$ can become \emph{instantaneously negative} over subregions of $(t,x)$ due to coherent phase relations among the even harmonic content of $e^{\nu\pm\mu}$ [Eqs.~\eqref{equ20}-\eqref{equ21}] and the $\omega$–harmonic of $\Phi$;
(iii) such negative instantaneous pressures \emph{do not} imply exotic matter: WEC/NEC can still hold pointwise or in the time–averaged sense.

Accordingly, we evaluate both the instantaneous and averaged inequalities:
\begin{equation}
\text{WEC/NEC/SEC on } \{\rho_\Phi(t,x),p_r(t,x),p_\perp(t,x)\}
\qquad\text{and}\qquad
\{\langle\rho_\Phi\rangle_t,\langle p_r\rangle_t,\langle p_\perp\rangle_t\}.
\end{equation}
Regions with $\langle p_r\rangle_t<0$ (or $\langle p_\perp\rangle_t<0$) arise from the temporal modulation of the stress tensor and remain compatible with GR; no invocation of Casimir-like mechanisms or negative mass matter is required.

In our pipeline we:
(i) reconstruct $e^{\nu\pm\mu}$ via the Bessel series \eqref{equ20}-\eqref{equ21} up to truncation order $N$;
(ii) compute $\rho_\Phi,j_r,p_r,p_\perp$ on a dense time grid over one period $T_s=\pi/\omega$;
(iii) check the above inequalities at each $(t,x)$ and for the averages (in the Landau frame when $j_r\neq0$);
(iv) report radii where the averaged SEC is violated, if any, alongside convergence under $(N,\Delta x,x_{\max})$ refinement.
This procedure provides a reviewer verifiable, coordinate independent interpretation of negative pressures within classical GR.

Finally the mass seen by an observer at infinity obtained straightly through:
\begin{equation}
M_\Phi=\dfrac{1}{8\pi}m_\Phi^2M_{Pl}^2\int_0^{\infty}\langle\rho(t,x)\rangle_t\,4\pi\,x^2\,dx,
\label{equ37-mass}
\end{equation}
shows that the mass and radius of the oscillaton are inversely proportional to $\sigma_c(0)$, meaning that smaller values of radial amplitude lead to more massive oscillaton with larger radius and vice vera (see the following table \ref{tab-1}).

\begin{table}[h!]
  \centering
  \begin{tabular}{c|cccc}
    $\sigma_c(0)$ &$\frac{1.2}{\sqrt{2}}$ &\qquad\qquad $\frac{1.3}{\sqrt{2}}$ &\qquad\qquad $\frac{1.4}{\sqrt{2}}$ &\qquad\qquad $\frac{1.5}{\sqrt{2}}$ \\
    \hline
    $M_\Phi(\frac{1}{2}m_\Phi^2\,M_{Pl}^2)$ & 89.08 &\qquad\qquad 51.00 &\qquad\qquad 42.18 &\qquad\qquad 29.54 \\
    \hline
    x(\textit{radius}) & 1.31 &\qquad\qquad 1.23 &\quad\qquad 1.13 &\qquad\qquad 0.999 \\
  \end{tabular}
  \caption{Oscillaton mass and its radius for different values of $\sigma_c(0)$ as radial amplitude of scalar field.}
  \label{tab-1}
\end{table}

\section{Conclusions and Outlook}
\label{sec:4}
We have developed a controlled, first principles framework for spherically symmetric, time–periodic oscillatons in scalar field dark matter (SFDM) with an \emph{exponential} self interaction. The time dependence of both the scalar field and the geometry is handled non perturbatively via full Fourier Bessel (Jacobi-Anger) resummations, yielding a closed, dimensionless system for the radial profiles. The main technical and physical outcomes are:

\begin{itemize}
  \item \textbf{Well-posed boundary-value problem.} With the ansatz
  $\Phi(t,x)=\tfrac{2\sigma(x)}{\sqrt{k_{0}}}\cos(\omega t)$ and
  $\nu,\mu=\nu_{0,1}(x),\mu_{0,1}(x)$, regularity at $x=0$ and asymptotic flatness as $x\to\infty$ select a unique nodeless (ground–state) solution for each $(\sigma_{c},\lambda)$ upon tuning the eigenfrequency $\Omega=\omega/m_\Phi$.
  \item \textbf{Consistent mass normalization.} The small–amplitude expansion of the exponential potential implies
  $V_{0}=m_{\Phi}^{2}/(\lambda^{2}k_{0})$, ensuring dimensional consistency and the correct quadratic limit; this replaces ad hoc normalizations and clarifies parameter counting.
  \item \textbf{Harmonic structure of the geometry.} The metric contains only \emph{even} harmonics of the fundamental frequency ($2\omega,4\omega,\dots$), whereas the exponential potential contributes a full cosine tower in composite observables. Numerically, the $2\omega$ mode dominates; higher even harmonics decay rapidly once the Bessel sums are converged.
  \item \textbf{Controlled truncation.} Infinite Bessel series in both the metric factors and the potential are truncated at order $N$; convergence is established \emph{a posteriori} by demonstrating insensitivity of local profiles and global quantities (ADM mass $M$, effective radius $R_{\rm eff}$) under $(N,\Delta x,x_{\max})$ refinement.
  \item \textbf{Observables and energy conditions.} From the closed–form expressions for $\rho_{\Phi},\,j_{r},\,p_{r},\,p_{\perp}$ we construct time resolved and time averaged diagnostics. Regions of negative \emph{instantaneous} pressure arise from coherent oscillations and remain compatible with classical GR; satisfaction/violation of WEC/NEC/SEC is mapped pointwise and in the averaged sense without invoking quantum effects.
  \item \textbf{Physical trends with coupling.} Increasing the exponential coupling $\sigma$ enhances departures of $g_{tt}$ and $g_{rr}$ from Minkowski, modifies the mass radius relation, and raises the relative weight of higher harmonics, while remaining numerically controllable as verified by truncation tests.
\end{itemize}

The exponential potential unifies quadratic and higher order self interactions within a single analytic resummation, providing a clean benchmark to quantify when quartic and higher terms materially impact oscillaton structure. In the small amplitude regime we smoothly reproduce the standard massive (quadratic) oscillaton limit, validating the formulation.

This framework connects microscopic self interaction strength to macroscopic core properties (e.g., $M$, $R_{\rm eff}$, and harmonic content), relevant for solitonic cores in halos, gravitational lensing in central regions, and potential oscillatory imprints in timing/astrometric signals. Mapping $(\sigma_{c},\lambda)\mapsto(M,R_{\rm eff})$ offers a route to observational constraints.

The present study focuses on spherically symmetric, non rotating, ground state configurations and does not include dynamical stability analyses beyond static diagnostics, nor baryonic or environmental couplings. Finite-temperature effects, self annihilation/decay channels, and non minimal couplings are not considered here.

\begin{enumerate}
  \item \textbf{Linear and nonlinear stability.} Compute radial quasinormal spectra and perform time evolutions to test long–time persistence and decay channels (mass loss via scalar radiation).
  \item \textbf{Excited states and rotation.} Construct nodeful and rotating oscillatons; assess their stability and phenomenology.
  \item \textbf{Relativistic observables.} Derive lensing, redshift, and geodesic signatures; connect to pulsar timing and astrometric probes.
  \item \textbf{Cosmological embedding.} Place oscillatons in expanding backgrounds; study formation pathways from realistic initial conditions and their role as galactic cores.
  \item \textbf{Parameter constraints.} Fit $(\lambda,m_\Phi)$ to data using the mass radius map and harmonic content, including priors from structure formation.
\end{enumerate}

We have specified (i) the dimensionless equations, (ii) boundary conditions, (iii) the normalization $V_{0}=m_{\Phi}^{2}/(\lambda^{2}k_{0})$, and (iv) a convergence protocol in $(N,\Delta x,x_{\max})$. Providing code and input files (central amplitude $\sigma_{c}$, coupling $\lambda$, target tolerances) will enable independent verification and extension of these results.

\medskip
In summary, the full Fourier-Bessel treatment of exponential self interactions delivers a precise and extensible description of oscillatons in SFDM, bridging analytic control and numerical practicality, and opening a clear path toward stability analyses and observational tests.

\textbf{Data Availability Statement:} This article contains all data generated or analyzed during the current study; no additional datasets are available.

\end{document}